\documentclass[twocolumn,aps,prc,superscriptaddress]{revtex4-2}
\usepackage{amssymb}
\usepackage{amsmath,bm}
\usepackage{graphicx}
\usepackage[normalem]{ulem}
\usepackage{multirow}
\usepackage[colorlinks,linkcolor=blue,urlcolor=blue,citecolor=blue]{hyperref}
\usepackage{lipsum}
\usepackage[usenames, dvipsnames]{xcolor}
\usepackage{tensor}
\usepackage{isotope}
\usepackage{amsmath}
\allowdisplaybreaks[4]
 
\renewcommand{\sout}{\bgroup \color{red} \ULdepth=-.5ex \ULset}
\newcommand{\com}[1]{{\sf\color[rgb]{0,0,1}{#1}}} 
 
\begin{document}

\title{Kinetic approach of light-nuclei production in intermediate-energy heavy-ion collisions}

\author{Rui Wang}
\email{wangrui@sinap.ac.cn}
\affiliation{Key Laboratory of Nuclear Physics and Ion-beam Application~(MOE), and Institute of Modern Physics, Fudan University, Shanghai $200433$, China}
\affiliation{Shanghai Institute of Applied Physics, Chinese Academy of Sciences, Shanghai $201800$, China}

\author{Yu-Gang Ma}
\email{mayugang@fudan.edu.cn}
\affiliation{Key Laboratory of Nuclear Physics and Ion-beam Application~(MOE), and Institute of Modern Physics, Fudan University, Shanghai $200433$, China}
\affiliation{Shanghai Research Center for Theoretical Nuclear Physics (NSFC), Fudan University, Shanghai $200438$, China}

\author{Lie-Wen Chen}
\email{lwchen@sjtu.edu.cn}
\affiliation{School of Physics and Astronomy, Shanghai Key Laboratory for Particle Physics and Cosmology, and Key Laboratory for Particle Astrophysics and Cosmology (MOE), Shanghai Jiao Tong University, Shanghai 200240, China}

\author{Che Ming Ko}
\email{ko@comp.tamu.edu}
\affiliation{Cyclotron Institute and Department of Physics and Astronomy, Texas A\&M University, College Station, Texas 77843, USA}

\author{Kai-Jia Sun}
\email{kjsun@tamu.edu}
\affiliation{Key Laboratory of Nuclear Physics and Ion-beam Application~(MOE), and Institute of Modern Physics, Fudan University, Shanghai $200433$, China}
\affiliation{Shanghai Research Center for Theoretical Nuclear Physics (NSFC), Fudan University, Shanghai $200438$, China}

\author{Zhen Zhang}
\email{zhangzh275@mail.sysu.edu.cn}
\affiliation{Sino-French Institute of Nuclear Engineering and Technology, Sun Yat-Sen University, Zhuhai 519082, China}

\date{\today}
\begin{abstract}
We develop a kinetic approach to the production of light nuclei up to mass number $A$ $\leqslant$ $4$ in intermediate-energy heavy-ion collisions by including them as dynamic degrees of freedom.
The conversions between nucleons and light nuclei during the collisions are incorporated dynamically via the breakup of light nuclei by a nucleon and their inverse reactions.
We also include the Mott effect on light nuclei, i.e., a light nucleus would no longer be bound
if the phase-space density of its surrounding nucleons is
too large.
With this kinetic approach, we obtain a reasonable description of the measured yields of light nuclei in central Au+Au collisions at energies of $0.25$ - $1.0A~\rm GeV$ by the FOPI collaboration. 
Our study also indicates that the observed enhancement of the $\alpha$-particle yield at low incident energies can be attributed to a weaker Mott effect on the $\alpha$-particle, which makes it more difficult to dissolve in nuclear medium, as a result of its much larger binding energy.
 \end{abstract}

\maketitle

Heavy-ion collisions from the Fermi energy to the GeV region has been extensively used to study the properties of nucleon-nucleon effective interactions and the nuclear equation of state~\cite{BarPR410,LBAPR464}.
Significant progresses have been achieved from studying in these collisions the nucleon and pion observables, such as the proton collective flow~\cite{DanSc298}, the neutron-to-proton spectral ratio~\cite{FamPRL97}, and the charged pion ratio~\cite{XZPRL102,EstPRL126}.
Since light nuclei are abundantly produced in heavy-ion collisions in this energy region, they are expected to have significant effects on the collision dynamics, which can then influence the nucleon and pion observables~\cite{OnoPPNP105}. Therefore, a reliable theoretical description of these collisions requires treating light nuclei on the same footing as nucleons and pions.
In this case, these collisions can also provide the possibility to study the in-medium properties of light nuclei and their relative abundances in warm nuclear matter~\cite{QLPRL108,HagPRL108,PaiPRL125}, which are known to have important implications in the dynamics of core-collapse supernovaeas, as well as the properties of compact stars and their mergers~\cite{SumPRC77,OerRMP89}.


Despite their great importance, light-nuclei observables in heavy-ion collisions have not recevied as much attention as the nucleon and pion observables, and they are also not explicitly included in most theoretical approaches for heavy-ion collisions.
Although there were attempts to describe light nuclei dynamically in transport models~\cite{DanNPA533,DanPRC46,KuhPRC63}, the $\alpha$-particle was not included in these studies.
Since then, new measurements of light nuclei up to mass number $A$ $\leqslant$ $4$ in heavy-ion collisions from the Fermi energy to the GeV region, especially for Au+Au collisions, have become available from the INDRA and FOPI collaborations~\cite{ReiNPA848,BouSmtr13}.
The measured data shows a significantly enhanced yield of $\alpha$-particles in collisions at low incident energies.  This surprising result has been suggested as an evidence for the Mott effect of light nuclei~\cite{HagPRL108}, i.e., a light nucleus would no longer be bound if the phase-space density of its surrounding
nucleons is too large~\cite{RopNPA379,RopNPA399}.
These new measurements call for a dynamical approach for these collisions that includes all light nuclei up to the $\alpha$-particle.

In the present study, based on the real-time many-body Green's-function formulism~\cite{Ram2007}, we develop a kinetic approach to intermediate-energy heavy-ion collisions by including both nucleon and light-nuclei ($A$ $\leqslant$ $4$) degrees of freedom.
Specifically, the production and dissociation of deuteron ($d$), triton ($t$), helium-3($h$) and $\alpha$-particle appear in this formulism as many-particle scatterings.
The Mott effects on these nuclei are also included explicitly by considering the nucleon phase-space density around them.
With this kinetic approach, we are able to reproduce the measured light-nuclei yields in central Au+Au collision at energies of $0.25$ - $1.0A~\rm GeV$ by the FOPI collaboration
We further show that the observed enhancement of $\alpha$-particle yield is a consequence of the Mott effect of light nuclei.


In the standard kinetic approach for heavy-ion collisions, such as the one based on the Boltzmann–Uehling-Uhlenbeck equation~\cite{BerPR160,WolPPNP125}, there is a truncation at two-particle scatterings and also only the nucleonic degrees of freedom are considered. To include light nuclei in the kinetic approach, one can resort to the real-time Green's-function formulism~\cite{Ram2007}, in which a light nucleus consisting of $A$ nucleons appears as a pole of the $A$-particle Green's function. The kinetic equations for light nuclei can then be derived by applying the Dyson equation in the vicinity of this pole~\cite{DanNPA533}. 
Including all light nuclei with $A$ $\leqslant$ $4$, we obtain the following coupled kinetic equations for the time evolution of their Wigner functions or phase-space distributions $f_{\tau}(\vec{r},\vec{p},t)$,
\begin{equation}
    (\partial_t + \vec{\nabla}_p\epsilon_{\tau}\cdot\vec{\nabla}_r - \vec{\nabla}_r\epsilon_{\tau}\cdot\vec{\nabla}_p)f_{\tau} = I_{\tau}^{\rm coll}[f_n,f_p,\cdots],
\label{E:KE}
\end{equation}
where $\tau$ represents $n$, $p$, $d$, $t$, $h$ and $\alpha$, as well as the pion ($\pi$) and $\Delta$-resonance.
In the above equation, $\epsilon_\tau[f_n,f_p,\cdots]$ is the single-particle energy of particle species $\tau$, and it is usually derived from a density functional.
The collision integral $I_{\tau}^{\rm coll}$ consists of a gain term ($<$) and a loss term ($>$),
\begin{equation}\label{coll}
I_{\tau}^{\rm coll} = K_{\tau}^{<}[f_n,f_p,\cdots](1\pm f_{\tau}) - K_{\tau}^{>}[f_n,f_p,\cdots]f_{\tau}\com{,}
\end{equation}
where the plus and minus signs are for bosons and fermions, respectively.
Both gain and loss terms contain contributions from various scattering channels, which can be obtained through the diagrammatic expansion of many-particle Green's function~\cite{DanNPA533}.  For light nuclei, we include the following nucleon-induced catalytic reactions $NNN\leftrightarrow Nd$, $NNNN\leftrightarrow Nt(h)$, $NNNNN\leftrightarrow N\alpha$, $NNt(h)\leftrightarrow N\alpha$, and the two-body inelastic channel $N\alpha\leftrightarrow dt(h)$. 
For the loosely bound deuteron, we do not include its production and absorption from $t$, $h$ and $\alpha$ breakup channels and their inverse reactions (e.g. $N\alpha\leftrightarrow NNNd$). For example, the $\alpha$-particle loss term $K_\alpha^{>}f_\alpha$ in Eq.(\ref{coll}) is expressed as
\begin{widetext}
\begin{equation}
\begin{split}
K_\alpha^{>}f_\alpha = &\frac{{\cal S}_{5'}f_\alpha}{2E_\alpha}\int \prod_{i=1'}^{5'}\frac{\text{d}\vec{p}_{i}}{(2\pi\hbar)^32E_{i}}\frac{\text{d}\vec{ p}_{N}}{(2\pi\hbar)^32E_N} \overline{|\mathcal{M}_{N\alpha\rightarrow NNNNN}|^2}g_N f_N \prod_{i=1'}^{5'} (1\pm f_i)  (2\pi)^4\delta^4(\sum_{i=1'}^{5'}p_i - p_N - p_\alpha)\\
+ &\frac{{\cal S}_{3'}f_\alpha}{2E_\alpha}\int \prod_{i=1'}^{3'}\frac{\text{d}\vec{p}_{i}}{(2\pi\hbar)^32E_{i}}\frac{\text{d}\vec{ p}_{N}}{(2\pi\hbar)^32E_N} \overline{|\mathcal{M}_{N\alpha\rightarrow NNt}|^2}g_N f_N\prod_{i=1'}^{3'} (1\pm f_i)  (2\pi)^4\delta^4(\sum_{i=1'}^{3'}p_i - p_N - p_\alpha) + t \rightarrow h\\
+ &\frac{{\cal S}_{2'}f_\alpha}{2E_\alpha}\int \prod_{i=1'}^{2'}\frac{\text{d}\vec{p}_{i}}{(2\pi\hbar)^32E_{i}}\frac{\text{d}\vec{ p}_{N}}{(2\pi\hbar)^32E_N} \overline{|\mathcal{M}_{N\alpha\rightarrow dt}|^2}g_N f_N\prod_{i=1'}^{2'} (1\pm f_i)  (2\pi)^4\delta^4(\sum_{i=1'}^{2'}p_i - p_N - p_\alpha) + t \rightarrow h.
\end{split}
\end{equation}
\end{widetext}
In the above, $1'$ - $5'$ denote final-state particles, $g_N$ is the spin degeneracy of nucleon, and ${\cal S}_{5'}$, ${\cal S}_{3'}$ and ${\cal S}_{2'}$ are symmetry factors that take into account possible identical particles in the final state of a reaction.

The transition amplitudes in the kinetic equations can be deduced from the experimental differential cross sections and the detailed balance relation. For catalytic reactions, this can be achieved using the impulse approximation. Under this approximation, the spin-averaged squared transition matrix element of a catalytic reaction is decomposed into the product of the internal momentum-space wave function of the light nucleus and the spin-averaged squared amplitude of the nucleon-nucleon elastic scattering amplitude $\overline{|M_{NN\rightarrow NN}|^2}$. As an example, the spin-averaged squared transition matrix element $\overline{|\mathcal{M}_{N\alpha\rightarrow NNNNN}|^2}$ for the reaction $N\alpha\to NNNNN$ is approximately written as
\begin{equation}
\begin{split}
&\overline{|\mathcal{M}_{N\alpha\rightarrow NNNNN}|^2}\\
&~~\approx F(\sqrt{s})\sum_{\rm spectator\atop nucleons}|\langle\vec{k}\vec{k}_\lambda\vec{k}_\mu|\phi_\alpha\rangle|^2\overline{|M_{NN\rightarrow NN}|^2},   
\end{split}
\label{E:IA}
\end{equation}
where $\vec{k}$, $\vec{k}_\lambda$ and $\vec{k}_\mu$ denote the three relative momenta between the constituent nucleons of the $\alpha$-particle. In the above, the summation runs over all combinations of spectator nucleons.
For simplicity, the internal wave functions of light nuclei are chosen to have a Gaussian form in the present study.

Since the cross section obtained from the factor $\sum_{\rm spectator\atop nucleons}|\langle\vec{k}\vec{k}_\lambda\vec{k}_\mu|\phi_\alpha\rangle|^2\overline{|M_{NN\rightarrow NN}|^2}$ in Eq.(\ref{E:IA}) for the reaction $N\alpha\to NNNNN$ may not agree with the measured one because of the neglect of elastic $N\alpha$  scattering and the possible inadequacy of the impulse approximation, a center-of-mass scattering energy $\sqrt{s}$ dependent factor $F(\sqrt{s})$ is introduced in Eq.(\ref{E:IA}) to account for these effects.
$F(\sqrt{s})$ can be determined from comparing the nucleon-nucleus scattering cross sections from the impulse approximation with those measured from experiments.
Because the nucleon-nucleus scattering at large incident energies is dominated by inelastic break-up reaction, we also require that the  factor $F(\sqrt{s})$, or more generally, the sum of $F(\sqrt{s})$ when there are many different outgoing channels, should approach $1.0$ as $\sqrt{s}$ increases.

\begin{figure}[htp]
\centering
\includegraphics[width=8.7CM]{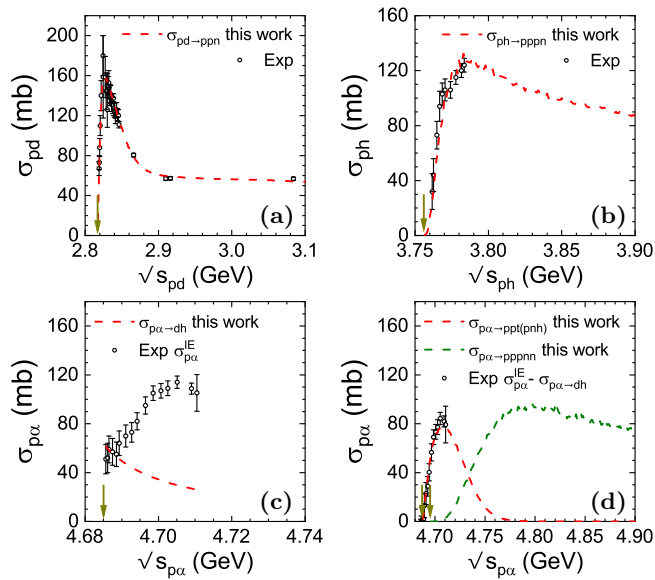}
\put(-150,135){\bfseries (a)}
\put(-25,135){\bfseries (b)}
\put(-150,27){\bfseries (c)}
\put(-25,27){\bfseries (d)}
\caption{Cross sections of inelastic (a) $pd$, (b) $ph$, (c) $p\alpha$ $\rightarrow$ $dh$, and (d) $p\alpha$ $\rightarrow$ $ppt(pnh)$ and $p\alpha$ $\rightarrow$ $pppnn$ from the impulse approximation (dashed lines).  The measured cross sections (open circles) are taken from Refs.~\cite{CarLNC8,SouPRC13} and references therein, with
$\sigma_{p\alpha}^{IE}$ in (c) being the measured inelastic $p\alpha$ cross section and after being subtracted by $\sigma_{p\alpha\rightarrow dh}$ in (d). 
The arrows denote the threshold of these reactions.}
\label{F:CS}
\end{figure}

We show in Fig.~\ref{F:CS} by dashed lines the cross sections of these break-up reactions used in the present kinetic approach, obtained by parametrizing the factor $F(\sqrt{s})$ for different nucleon-nucleus scatterings to reproduce the corresponding cross sections measured in experiments.
Fig.~\ref{F:CS}(a) and Fig.~\ref{F:CS}(b) show, respectively, the break-up cross section of $pd$ and $ph$.
For the break-up of $p\alpha$, there are three different final-state channels of $dh$, $ppt(pnh)$ and $pppnn$.
The $p\alpha$ $\rightarrow$ $dh$ channel is included to account for the cross section below the $p\alpha$ $\rightarrow$ $ppt(pnh)$ threshold, whose cross section is shown in Fig.~\ref{F:CS}(c).
The cross section $\sigma_{p\alpha\rightarrow dh}$ is deduced from the measured cross section of the reaction $dt$ $\rightarrow$ $n\alpha$~\cite{BamPR107} using the detailed balance relation.
Apart from $p\alpha$ $\rightarrow$ $dh$, the total inelastic $p\alpha$ cross section is largely exhausted by $p\alpha$ $\rightarrow$ $ppt(pnh)$ at small $\sqrt{s_{p\alpha}}$~[red line in Fig.~\ref{F:CS}(d)] and by $p\alpha$ $\rightarrow$ $pppnn$ at large $\sqrt{s_{p\alpha}}$~[olive line in Fig.~\ref{F:CS}(d)].
The above assumption for the branching ratios of inelastic $p\alpha$ scattering is based on the argument that a proton with higher incident energy makes it easier for the $\alpha$ particle to fully breakup.



One of the important features of light nuclei in a nuclear medium is the Mott effect on their binding energies, i.e., they would no longer be bound if the phase-space density of their surrounding nucleons is too large.
To include the Mott effect on a light nucleus, one should in principle solve an in-medium Schrodinger equation, which takes into account the Pauli-blocking effect, for the light nucleus moving with a momentum $\vec P$ in the nuclear medium~\cite{RopNPA379,RopNPA399}.
Because of the Pauli blocking of the constituent nucleons in a light nucleus due to the nucleons in nuclear medium, the resulting binding energy $E_{\rm B}(\vec{P})$ is expected to decrease with increasing nucleon phase-space density in the nuclear medium.
For a sufficiently large nucleon phase-space density around the light nucleus in the nuclear medium, $E_{\rm B}(\vec{P})$ would vanish, and the light nucleus would no longer be bound.
This criterion for the existence of light nuclei can be effectively implemented in the kinetic approach by introducing a phase-space cutoff in the collision integral for their production. Specifically, $A$ free nucleons of total momentum $\vec P$ in a nuclear medium are allowed to form a nucleus of mass number $A$ only if the average nucleon phase-space density of the medium around the light nucleus is less than a cutoff value $f_A^{\rm cut}$~\cite{DanNPA533}, i.e., 
\begin{eqnarray}
\langle f_N \rangle_A \equiv \int f_N\bigg(\frac{\vec{P}}{A} + \vec{p}\bigg)\rho_A(\vec{p}){\rm d}\vec{p} \leqslant f_A^{\rm cut},
\label{E:fcut}
\end{eqnarray}
where $\rho_A(\vec p)$ denotes the nucleon momentum distribution inside the light nucleus (related to its internal wave function), and $f_N$ is the nucleon phase-space distribution in the medium.

\begin{figure}[htb]
\centering
\includegraphics[width=6.0cm]{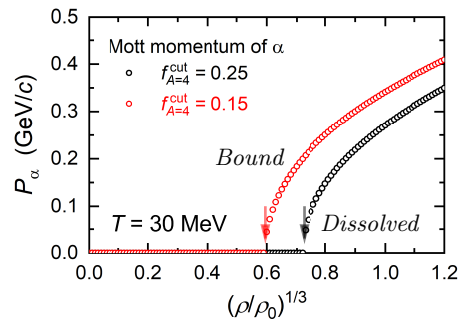}
\put(-93,60){\bfseries \it{Bound}}
\put(-53,35){\bfseries \it{Dissolved}}
\caption{Density dependence of the Mott momentum of $\alpha$-particle in nuclear matter at temperature $T=30~\rm MeV$, with $\rho_0=0.16~{\rm fm}^{-3}$ denoting the normal nuclear matter density. The arrows represent their corresponding Mott densities. The results are obtained with $f_{A=4}^{\rm cut}$ $=$ $0.25$ or $0.15$.}
\label{F:P_Mt}
\end{figure}

For nuclear matter in thermal equilibrium with $f_N$ given by the Fermi distribution, $E_{\rm B}(\vec{P})$ decreases with decreasing $|P|$ and vanishes below a critical momentum called the Mott momentum $P_{\rm Mott}$.
It has been shown that the density dependence of $P_{\rm Mott}$ obtained from Eq.~(\ref{E:fcut}) at a given temperature $T$ for deuteron and triton are consistent with those from the $t$-matrix approach~\cite{KuhPRC63}, and the preferred value of $f_A^{\rm cut}$ shows little temperature dependence~\cite{KuhPRC63}. In Fig.~\ref{F:P_Mt}, we show the density dependence of the Mott momentum of $\alpha$-particle obtained for two different values of 0.25 and 0.15 for $f_{A=4}^{\rm cut}$ in nuclear matter at $T$ $=$ $30~\rm MeV$, which is the typical temperature reached in intermediate-energy heavy-ion collisions.
The Mott density of a light nucleus is then given by the maximum density at which a light nucleus of zero momentum can still be bound, as indicated by the arrows in the figure for the $\alpha$-particle. 
The cut-off parameters $f_{A=2}^{\rm cut}$, $f_{A=3}^{\rm cut}$ and $f_{A=4}^{\rm cut}$ can be considered as a surrogate for the strength of the Mott effects on deuteron, triton or helium-3, and $\alpha$-particle, respectively.
A smaller $f_{A}^{\rm cut}$ corresponds to a stronger Mott effect and a larger $P_{\rm Mott}$. 
For the implications of the values of $f_A^{\rm cut}$ on the in-medium properties of light nuclei in nuclear matter, we leave it to a future study, and in the present study we treat them only as parameters for reproducing measured yields of light nuclei in intermediate-energy heavy-ion collisions.

We solve the kinetic equations by employing the test particle ansatz~\cite{WonPRC25}, which approximates $f_\tau$ in terms of a large number of $\delta$-functions, i.e., $f_{\tau}(\vec{r},\vec{p})$ $\approx$ $\frac{(2\pi\hbar)^3}{g_{\tau} N_{\rm E}}\sum^{N_{\tau}N_{\rm E}}_{i=1}\delta(\vec{r}_i - \vec{ r})\delta(\vec{p}_i - \vec{p})$, where
$g_{\tau}$ and $N_E$ denote, respectively, the spin degeneracy of particle species $\tau$ and the number of test particle or ensemble used in solving the kinetic equations.
To ensure the convergence of numerical results, a sufficiently large $N_{\rm E}$ is used.
To improve the numerical accuracy, we further adopt the lattice Hamiltonian method~\cite{LenPRC39,WRPRC99} to treat the drift terms on the left-hand side of Eq.~(\ref{E:KE}). 
As to the single-particle energy $\epsilon_\tau$ in Eq.~(\ref{E:KE}), we use the one derived from the Skyrme pseudo-potential~\cite{RaiPRC83,WRPRC98}. For the collision integral on the right-hand side of Eq.~(\ref{E:KE}), it is treated by the stochastic method~\cite{DanNPA533,WRPLB807}, in which the scattering probability of initial-state (test-)particles within a time interval is calculated directly from the loss term $K_{\tau}^{>}f_{\tau}$.

In the present study, we apply the above kinetic approach to central Au+Au collisions at the incident energy from $E_{\rm beam}$ $=$ $0.25$ to $1.0A~\rm GeV$. Besides elastic scatterings and the many-particle scatterings related to light-nuclei production and dissociation, we include in the kinetic approach also scatterings related to $\Delta$-resonances and pions, i.e. $NN$ $\leftrightarrow$ $N\Delta$ and $\Delta$ $\leftrightarrow$ $N\pi$~\cite{OnoPRC100}.
Since in heavy-ion collisions in this energy region, nucleons still dominate over pions, we neglect the production and dissociate of light nuclei with the pion as the catalyzer~\cite{OliPRC99,SKJX2022}.

\begin{figure}[ht]
\centering
\includegraphics[width=6.5cm]{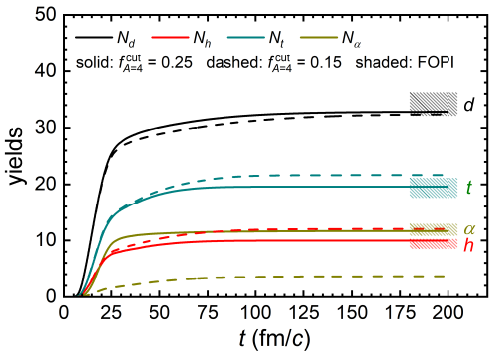}
\caption{Light-nuclei yields as functions of elapsed collision time in central Au+Au collision at $0.4A~\rm GeV$ from the kinetic approach with $f_{A=2}^{\rm cut}$ $=$ $0.11$, $f_{A=3}^{\rm cut}$ $=$ $0.16$, and two different $f_{A=4}^{\rm cut}$ $=$ $0.25$ and $0.15$.
The shaded areas represent the data measured by the FOPI Collaboration~\cite{ReiNPA848}.}
\label{F:400}
\end{figure}

We first show in Fig.~\ref{F:400} the time evolution of light-nuclei yields in central Au+Au collision at $0.4A~\rm GeV$ from our kinetic approach, with $f_{A=2}^{\rm cut}$ $=$ $0.11$ for deuteron, $f_{A=3}^{\rm cut}$ $=$ $0.16$ for triton and helium-3, and two different $f_{A=4}^{\rm cut}$ $=$ $0.25$ and $0.15$ for $\alpha$-particle. It is seen that decreasing $f_{A=4}^{\rm cut}$ significantly reduces the $\alpha$-particle yield. We also notice from the figure that the number of light nuclei increased significantly in the early phase of the time evolution, which corresponds to the compressing stage of the collision, because of the enhanced production rate of light nuclei in dense nuclear matter.
Since light nuclei have been already abundantly produced during the early compression stage of intermediate-energy heavy ion collisions, it is important to include them dynamically throughout the collisions, rather than to introduce them merely at the kinetic freeze-out of the collisions like in the coalescence model.

\begin{figure}[ht]
\centering
\includegraphics[width=8.5cm]{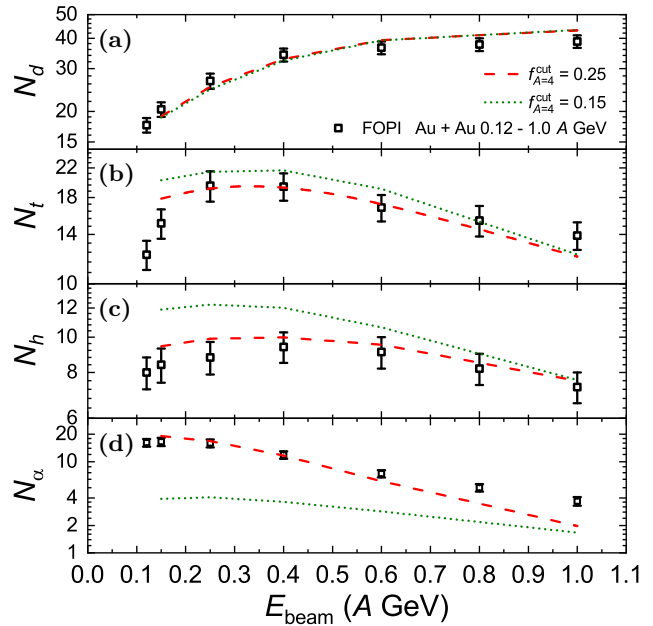}
\put(-207,221){\bfseries (a)}
\put(-207,170){\bfseries (b)}
\put(-207,119){\bfseries (c)}
\put(-207,68){\bfseries (d)}
\caption{Incident-energy dependence of light-nuclei yields from the kinetic approach with $f_{A=2}^{\rm cut}$ $=$ $0.11$, $f_{A=3}^{\rm cut}$ $=$ $0.16$ and $f_{A=4}^{\rm cut}$ $=$ $0.25$. The results for a smaller $f_{A=4}^{\rm cut}$ $=$ $0.15$ are also included for comparison. The experimental data are from the FOPI Collaboration~\cite{ReiNPA848}.}
\label{F:yld}
\end{figure}

In Fig.~\ref{F:yld}, we show the beam-energy dependence of light-nuclei yields in central Au+Au collisions from the kinetic approach.
They are obtained with the Mott effect of light nuclei properly incorporated by choosing appropriate values for the cutoff parameters $f^{\rm cut}_A$.
Due to the tight binding of $\alpha$-particle in free space, it is more difficult for $\alpha$-particle to dissolve in nuclear medium than deuteron, triton and helium-3, resulting in a weaker Mott effect and a smaller $P_{\rm Mott}/A$ for $\alpha$-particles in nuclear medium.
This is the same argument used in the calculation of the properties of nuclear matter with light nuclei from the quantum statistical approach and the generalized relativistic mean-field model~\cite{TypPRC81,ZZWPRC95}. It is also consistent with the larger Mott density of $\alpha$-particle than that of the deuteron, triton and helium-3 deduced from experiments~\cite{HagPRL108}.
This explains the larger value we have used for $f_{A=4}^{\rm cut}$ than those for $f_{A=2}^{\rm cut}$ and $f_{A=3}^{\rm cut}$.

It is seen in Fig.~\ref{F:yld} that the present kinetic approach with $f_{A=2}^{\rm cut}$ $=$ $0.11$, $f_{A=3}^{\rm cut}$ $=$ $0.16$ and $f_{A=4}^{\rm cut}$ $=$ $0.25$ reproduces reasonably the measured light-nuclei yields in central Au+Au collisions~\cite{ReiNPA848} for a wide range of incident energies, especially for the large $\alpha$-particle yield at lower incident energies.
As shown in Fig.~\ref{F:yld}, the measured yield of $\alpha$-particles at low incident energies surpasses that of helium-3, which is in sharp contrast to the prediction from the thermal model, which gives a decreasing yield with increasing mass number of light nuclei.
If we had used a smaller  $f_{A=4}^{\rm cut}$ $=$ $0.15$~(while fixing $f_{A=2}^{\rm cut}$ and $f_{A=3}^{\rm cut}$), which corresponds to a stronger Mott effect and a larger Mott momentum for the  $\alpha$-particle, the $\alpha$-particle yield would significantly reduced as shown in Fig.~\ref{F:yld}(d).
Our result thus indicates that the observed enhancement of the $\alpha$-particle yield in lower-energy collisions can be attributed to the weaker Mott effect on $\alpha$-particle than that on deuteron, triton and helium-3, as a result of its
much larger binding energy.

 


In summary, to provide a dynamical description of light-nuclei production in intermediate-energy heavy-ion collisions, we have included the light-nuclei degrees of freedom with $A$ $\leqslant$ $4$ into the kinetic approach.
The breakup of light nuclei by nucleons and their inverse reactions are included to account for the conversion between nucleons and light-nuclei during the collisions.
The Mott effects of light nuclei are also included by considering the nucleon phase-space density $\langle f_N \rangle$ around them, and a light nucleus can exist only if $\langle f_N \rangle$ is less than a cutoff value $f_A^{\rm cut}$.
With appropriate values of $f_A^{\rm cut}$ for different species of light nuclei, the present kinetic approach has reasonably reproduced the yields of light nuclei in central Au+Au collisions at incident energies from $0.25A~\rm GeV $ to $1.0A~\rm GeV$ measured by the FOPI Collaboration.
Our study has clearly demonstrated that the observed enhancement of the $\alpha$-particle yield compared with that of helium-3 at low incident energies is a consequence of the Mott effect of light nuclei. 
Therefore, studying the light-nuclei yields in intermediate-energy heavy-ion collisions allows one to determine the cutoff parameters $f^{\rm cut}_A$  and thus the strength of their Mott effect.
The implications of the preferred values of $f_A^{\rm cut}$ obtained in this work on the medium properties of light nuclei in warm nuclear matter will be reported in a forthcoming study.

The present kinetic approach can be further used to study phenomena related to light nuclei in nuclear reactions, such as the iso-scaling in intermediate-energy heavy-ion collisions~\cite{LeeEPJA58} and the effect of the $\alpha$-clusters formed on the surface of heavy nuclei~\cite{TanSc371}, as well as the role of light nuclei in core-collapse supernovae, compact stars and their merger~\cite{SumPRC77,OerRMP89}.  
Since the nuclear matter produced in heavy-ion collisions around the Fermi energy
could undergo the spinodal transition~\cite{ChoPR389,WRPRR2,LCNST33}, which would lead to the production of heavy fragments with mass number $A$ $\geqslant$ $5$.  To describe the dynamics of these heavy fragments requires the extension of the standard kinetic approach, as used in the present study, to include the fluctuations of nucleon phase-space distributions or Wigner functions~\cite{ColNPA642}. A possible and worthwhile further development of the present approach is to include such fluctuations in the present kinetic approach, so that low-energy nuclear reactions can also be properly described. These studies will be pursued in the future.

~

We thank Chen Zhong for setting up and maintaining the GPU server. This work was supported in part by the National Natural Science Foundation of China under contract Nos. $11890714$, $12147101$, $12235010$ and $11625521$, the National Key Research and Development Program of China under Grant Nos. $2018$YFE$0104600$ and $2022$YFA$1602303$, the National SKA Program of China No. $2020$SKA$0120300$, the Guangdong Major Project of Basic and Applied Basic Research No. $2020$B$0301030008$, and the U.S. Department of Energy under Award No. DE-SC$0015266$.

%


\begin{thebibliography}{42}%
\makeatletter
\providecommand \@ifxundefined [1]{%
 \@ifx{#1\undefined}
}%
\providecommand \@ifnum [1]{%
 \ifnum #1\expandafter \@firstoftwo
 \else \expandafter \@secondoftwo
 \fi
}%
\providecommand \@ifx [1]{%
 \ifx #1\expandafter \@firstoftwo
 \else \expandafter \@secondoftwo
 \fi
}%
\providecommand \natexlab [1]{#1}%
\providecommand \enquote  [1]{``#1''}%
\providecommand \bibnamefont  [1]{#1}%
\providecommand \bibfnamefont [1]{#1}%
\providecommand \citenamefont [1]{#1}%
\providecommand \href@noop [0]{\@secondoftwo}%
\providecommand \href [0]{\begingroup \@sanitize@url \@href}%
\providecommand \@href[1]{\@@startlink{#1}\@@href}%
\providecommand \@@href[1]{\endgroup#1\@@endlink}%
\providecommand \@sanitize@url [0]{\catcode `\\12\catcode `\$12\catcode
  `\&12\catcode `\#12\catcode `\^12\catcode `\_12\catcode `\%12\relax}%
\providecommand \@@startlink[1]{}%
\providecommand \@@endlink[0]{}%
\providecommand \url  [0]{\begingroup\@sanitize@url \@url }%
\providecommand \@url [1]{\endgroup\@href {#1}{\urlprefix }}%
\providecommand \urlprefix  [0]{URL }%
\providecommand \Eprint [0]{\href }%
\providecommand \doibase [0]{https://doi.org/}%
\providecommand \selectlanguage [0]{\@gobble}%
\providecommand \bibinfo  [0]{\@secondoftwo}%
\providecommand \bibfield  [0]{\@secondoftwo}%
\providecommand \translation [1]{[#1]}%
\providecommand \BibitemOpen [0]{}%
\providecommand \bibitemStop [0]{}%
\providecommand \bibitemNoStop [0]{.\EOS\space}%
\providecommand \EOS [0]{\spacefactor3000\relax}%
\providecommand \BibitemShut  [1]{\csname bibitem#1\endcsname}%
\let\auto@bib@innerbib\@empty
\bibitem [{\citenamefont {Baran}\ \emph {et~al.}(2005)\citenamefont {Baran},
  \citenamefont {Colonna}, \citenamefont {Greco},\ and\ \citenamefont
  {Di~Toro}}]{BarPR410}%
  \BibitemOpen
  \bibfield  {author} {\bibinfo {author} {\bibfnamefont {V.}~\bibnamefont
  {Baran}}, \bibinfo {author} {\bibfnamefont {M.}~\bibnamefont {Colonna}},
  \bibinfo {author} {\bibfnamefont {V.}~\bibnamefont {Greco}},\ and\ \bibinfo
  {author} {\bibfnamefont {M.}~\bibnamefont {Di~Toro}},\ }\bibfield  {title}
  {\bibinfo {title} {Reaction dynamics with exotic nuclei},\ }\href
  {https://doi.org/10.1016/j.physrep.2004.12.004} {\bibfield  {journal}
  {\bibinfo  {journal} {Phys. Rep.}\ }\textbf {\bibinfo {volume} {410}},\
  \bibinfo {pages} {335} (\bibinfo {year} {2005})}\BibitemShut {NoStop}%
\bibitem [{\citenamefont {Li}\ \emph {et~al.}(2008)\citenamefont {Li},
  \citenamefont {Chen},\ and\ \citenamefont {Ko}}]{LBAPR464}%
  \BibitemOpen
  \bibfield  {author} {\bibinfo {author} {\bibfnamefont {B.-A.}\ \bibnamefont
  {Li}}, \bibinfo {author} {\bibfnamefont {L.-W.}\ \bibnamefont {Chen}},\ and\
  \bibinfo {author} {\bibfnamefont {C.~M.}\ \bibnamefont {Ko}},\ }\bibfield
  {title} {\bibinfo {title} {Recent progress and new challenges in isospin
  physics with heavy-ion reactions},\ }\href
  {https://doi.org/10.1016/j.physrep.2008.04.005} {\bibfield  {journal}
  {\bibinfo  {journal} {Phys. Rep.}\ }\textbf {\bibinfo {volume} {464}},\
  \bibinfo {pages} {113} (\bibinfo {year} {2008})}\BibitemShut {NoStop}%
\bibitem [{\citenamefont {Danielewicz}\ \emph {et~al.}(2002)\citenamefont
  {Danielewicz}, \citenamefont {Lacey},\ and\ \citenamefont
  {Lynch}}]{DanSc298}%
  \BibitemOpen
  \bibfield  {author} {\bibinfo {author} {\bibfnamefont {P.}~\bibnamefont
  {Danielewicz}}, \bibinfo {author} {\bibfnamefont {R.}~\bibnamefont {Lacey}},\
  and\ \bibinfo {author} {\bibfnamefont {W.~G.}\ \bibnamefont {Lynch}},\
  }\bibfield  {title} {\bibinfo {title} {Determination of the {{Equation}} of
  {{State}} of {{Dense Matter}}},\ }\href
  {https://doi.org/10.1126/science.1078070} {\bibfield  {journal} {\bibinfo
  {journal} {Science}\ }\textbf {\bibinfo {volume} {298}},\ \bibinfo {pages}
  {1592} (\bibinfo {year} {2002})}\BibitemShut {NoStop}%
\bibitem [{\citenamefont {Famiano}\ \emph {et~al.}(2006)\citenamefont
  {Famiano}, \citenamefont {Liu}, \citenamefont {Lynch}, \citenamefont {Mocko},
  \citenamefont {Rogers}, \citenamefont {Tsang}, \citenamefont {Wallace},
  \citenamefont {Charity}, \citenamefont {Komarov}, \citenamefont {Sarantites},
  \citenamefont {Sobotka},\ and\ \citenamefont {Verde}}]{FamPRL97}%
  \BibitemOpen
  \bibfield  {author} {\bibinfo {author} {\bibfnamefont {M.~A.}\ \bibnamefont
  {Famiano}}, \bibinfo {author} {\bibfnamefont {T.}~\bibnamefont {Liu}},
  \bibinfo {author} {\bibfnamefont {W.~G.}\ \bibnamefont {Lynch}}, \bibinfo
  {author} {\bibfnamefont {M.}~\bibnamefont {Mocko}}, \bibinfo {author}
  {\bibfnamefont {A.~M.}\ \bibnamefont {Rogers}}, \bibinfo {author}
  {\bibfnamefont {M.~B.}\ \bibnamefont {Tsang}}, \bibinfo {author}
  {\bibfnamefont {M.~S.}\ \bibnamefont {Wallace}}, \bibinfo {author}
  {\bibfnamefont {R.~J.}\ \bibnamefont {Charity}}, \bibinfo {author}
  {\bibfnamefont {S.}~\bibnamefont {Komarov}}, \bibinfo {author} {\bibfnamefont
  {D.~G.}\ \bibnamefont {Sarantites}}, \bibinfo {author} {\bibfnamefont
  {L.~G.}\ \bibnamefont {Sobotka}},\ and\ \bibinfo {author} {\bibfnamefont
  {G.}~\bibnamefont {Verde}},\ }\bibfield  {title} {\bibinfo {title} {Neutron
  and {{Proton Transverse Emission Ratio Measurements}} and the {{Density
  Dependence}} of the {{Asymmetry Term}} of the {{Nuclear Equation}} of
  {{State}}},\ }\href {https://doi.org/10.1103/PhysRevLett.97.052701}
  {\bibfield  {journal} {\bibinfo  {journal} {Phys. Rev. Lett.}\ }\textbf
  {\bibinfo {volume} {97}},\ \bibinfo {pages} {052701} (\bibinfo {year}
  {2006})}\BibitemShut {NoStop}%
\bibitem [{\citenamefont {Xiao}\ \emph {et~al.}(2009)\citenamefont {Xiao},
  \citenamefont {Li}, \citenamefont {Chen}, \citenamefont {Yong},\ and\
  \citenamefont {Zhang}}]{XZPRL102}%
  \BibitemOpen
  \bibfield  {author} {\bibinfo {author} {\bibfnamefont {Z.}~\bibnamefont
  {Xiao}}, \bibinfo {author} {\bibfnamefont {B.-A.}\ \bibnamefont {Li}},
  \bibinfo {author} {\bibfnamefont {L.-W.}\ \bibnamefont {Chen}}, \bibinfo
  {author} {\bibfnamefont {G.-C.}\ \bibnamefont {Yong}},\ and\ \bibinfo
  {author} {\bibfnamefont {M.}~\bibnamefont {Zhang}},\ }\bibfield  {title}
  {\bibinfo {title} {Circumstantial {{Evidence}} for a {{Soft Nuclear Symmetry
  Energy}} at {{Suprasaturation Densities}}},\ }\href
  {https://doi.org/10.1103/PhysRevLett.102.062502} {\bibfield  {journal}
  {\bibinfo  {journal} {Phys. Rev. Lett.}\ }\textbf {\bibinfo {volume} {102}},\
  \bibinfo {pages} {062502} (\bibinfo {year} {2009})}\BibitemShut {NoStop}%
\bibitem [{\citenamefont {{S{$\pi$}RIT Collaboration}}\ and\ \citenamefont
  {Cozma}(2021)}]{EstPRL126}%
  \BibitemOpen
  \bibfield  {author} {\bibinfo {author} {\bibnamefont {{S{$\pi$}RIT
  Collaboration}}}\ and\ \bibinfo {author} {\bibfnamefont {M.~D.}\ \bibnamefont
  {Cozma}},\ }\bibfield  {title} {\bibinfo {title} {Probing the {{Symmetry
  Energy}} with the {{Spectral Pion Ratio}}},\ }\href
  {https://doi.org/10.1103/PhysRevLett.126.162701} {\bibfield  {journal}
  {\bibinfo  {journal} {Phys. Rev. Lett.}\ }\textbf {\bibinfo {volume} {126}},\
  \bibinfo {pages} {162701} (\bibinfo {year} {2021})}\BibitemShut {NoStop}%
\bibitem [{\citenamefont {Ono}(2019)}]{OnoPPNP105}%
  \BibitemOpen
  \bibfield  {author} {\bibinfo {author} {\bibfnamefont {A.}~\bibnamefont
  {Ono}},\ }\bibfield  {title} {\bibinfo {title} {Dynamics of clusters and
  fragments in heavy-ion collisions},\ }\href
  {https://doi.org/10.1016/j.ppnp.2018.11.001} {\bibfield  {journal} {\bibinfo
  {journal} {Prog. Part. Nucl. Phys.}\ }\textbf {\bibinfo {volume} {105}},\
  \bibinfo {pages} {139} (\bibinfo {year} {2019})}\BibitemShut {NoStop}%
\bibitem [{\citenamefont {Qin}\ \emph {et~al.}(2012)\citenamefont {Qin},
  \citenamefont {Hagel}, \citenamefont {Wada}, \citenamefont {Natowitz},
  \citenamefont {Shlomo}, \citenamefont {Bonasera}, \citenamefont {R{\"o}pke},
  \citenamefont {Typel}, \citenamefont {Chen}, \citenamefont {Huang},
  \citenamefont {Wang}, \citenamefont {Zheng}, \citenamefont {Kowalski},
  \citenamefont {Barbui}, \citenamefont {Rodrigues}, \citenamefont {Schmidt},
  \citenamefont {Fabris}, \citenamefont {Lunardon}, \citenamefont {Moretto},
  \citenamefont {Nebbia}, \citenamefont {Pesente}, \citenamefont {Rizzi},
  \citenamefont {Viesti}, \citenamefont {Cinausero}, \citenamefont {Prete},
  \citenamefont {Keutgen}, \citenamefont {El~Masri}, \citenamefont {Majka},\
  and\ \citenamefont {Ma}}]{QLPRL108}%
  \BibitemOpen
  \bibfield  {author} {\bibinfo {author} {\bibfnamefont {L.}~\bibnamefont
  {Qin}}, \bibinfo {author} {\bibfnamefont {K.}~\bibnamefont {Hagel}}, \bibinfo
  {author} {\bibfnamefont {R.}~\bibnamefont {Wada}}, \bibinfo {author}
  {\bibfnamefont {J.~B.}\ \bibnamefont {Natowitz}}, \bibinfo {author}
  {\bibfnamefont {S.}~\bibnamefont {Shlomo}}, \bibinfo {author} {\bibfnamefont
  {A.}~\bibnamefont {Bonasera}}, \bibinfo {author} {\bibfnamefont
  {G.}~\bibnamefont {R{\"o}pke}}, \bibinfo {author} {\bibfnamefont
  {S.}~\bibnamefont {Typel}}, \bibinfo {author} {\bibfnamefont
  {Z.}~\bibnamefont {Chen}}, \bibinfo {author} {\bibfnamefont {M.}~\bibnamefont
  {Huang}}, \bibinfo {author} {\bibfnamefont {J.}~\bibnamefont {Wang}},
  \bibinfo {author} {\bibfnamefont {H.}~\bibnamefont {Zheng}}, \bibinfo
  {author} {\bibfnamefont {S.}~\bibnamefont {Kowalski}}, \bibinfo {author}
  {\bibfnamefont {M.}~\bibnamefont {Barbui}}, \bibinfo {author} {\bibfnamefont
  {M.~R.~D.}\ \bibnamefont {Rodrigues}}, \bibinfo {author} {\bibfnamefont
  {K.}~\bibnamefont {Schmidt}}, \bibinfo {author} {\bibfnamefont
  {D.}~\bibnamefont {Fabris}}, \bibinfo {author} {\bibfnamefont
  {M.}~\bibnamefont {Lunardon}}, \bibinfo {author} {\bibfnamefont
  {S.}~\bibnamefont {Moretto}}, \bibinfo {author} {\bibfnamefont
  {G.}~\bibnamefont {Nebbia}}, \bibinfo {author} {\bibfnamefont
  {S.}~\bibnamefont {Pesente}}, \bibinfo {author} {\bibfnamefont
  {V.}~\bibnamefont {Rizzi}}, \bibinfo {author} {\bibfnamefont
  {G.}~\bibnamefont {Viesti}}, \bibinfo {author} {\bibfnamefont
  {M.}~\bibnamefont {Cinausero}}, \bibinfo {author} {\bibfnamefont
  {G.}~\bibnamefont {Prete}}, \bibinfo {author} {\bibfnamefont
  {T.}~\bibnamefont {Keutgen}}, \bibinfo {author} {\bibfnamefont
  {Y.}~\bibnamefont {El~Masri}}, \bibinfo {author} {\bibfnamefont
  {Z.}~\bibnamefont {Majka}},\ and\ \bibinfo {author} {\bibfnamefont {Y.~G.}\
  \bibnamefont {Ma}},\ }\bibfield  {title} {\bibinfo {title} {Laboratory
  {{Tests}} of {{Low Density Astrophysical Nuclear Equations}} of {{State}}},\
  }\href {https://doi.org/10.1103/PhysRevLett.108.172701} {\bibfield  {journal}
  {\bibinfo  {journal} {Phys. Rev. Lett.}\ }\textbf {\bibinfo {volume} {108}},\
  \bibinfo {pages} {172701} (\bibinfo {year} {2012})}\BibitemShut {NoStop}%
\bibitem [{\citenamefont {Hagel}\ \emph {et~al.}(2012)\citenamefont {Hagel},
  \citenamefont {Wada}, \citenamefont {Qin}, \citenamefont {Natowitz},
  \citenamefont {Shlomo}, \citenamefont {Bonasera}, \citenamefont {R{\"o}pke},
  \citenamefont {Typel}, \citenamefont {Chen}, \citenamefont {Huang},
  \citenamefont {Wang}, \citenamefont {Zheng}, \citenamefont {Kowalski},
  \citenamefont {Bottosso}, \citenamefont {Barbui}, \citenamefont {Rodrigues},
  \citenamefont {Schmidt}, \citenamefont {Fabris}, \citenamefont {Lunardon},
  \citenamefont {Moretto}, \citenamefont {Nebbia}, \citenamefont {Pesente},
  \citenamefont {Rizzi}, \citenamefont {Viesti}, \citenamefont {Cinausero},
  \citenamefont {Prete}, \citenamefont {Keutgen}, \citenamefont {El~Masri},\
  and\ \citenamefont {Majka}}]{HagPRL108}%
  \BibitemOpen
  \bibfield  {author} {\bibinfo {author} {\bibfnamefont {K.}~\bibnamefont
  {Hagel}}, \bibinfo {author} {\bibfnamefont {R.}~\bibnamefont {Wada}},
  \bibinfo {author} {\bibfnamefont {L.}~\bibnamefont {Qin}}, \bibinfo {author}
  {\bibfnamefont {J.~B.}\ \bibnamefont {Natowitz}}, \bibinfo {author}
  {\bibfnamefont {S.}~\bibnamefont {Shlomo}}, \bibinfo {author} {\bibfnamefont
  {A.}~\bibnamefont {Bonasera}}, \bibinfo {author} {\bibfnamefont
  {G.}~\bibnamefont {R{\"o}pke}}, \bibinfo {author} {\bibfnamefont
  {S.}~\bibnamefont {Typel}}, \bibinfo {author} {\bibfnamefont
  {Z.}~\bibnamefont {Chen}}, \bibinfo {author} {\bibfnamefont {M.}~\bibnamefont
  {Huang}}, \bibinfo {author} {\bibfnamefont {J.}~\bibnamefont {Wang}},
  \bibinfo {author} {\bibfnamefont {H.}~\bibnamefont {Zheng}}, \bibinfo
  {author} {\bibfnamefont {S.}~\bibnamefont {Kowalski}}, \bibinfo {author}
  {\bibfnamefont {C.}~\bibnamefont {Bottosso}}, \bibinfo {author}
  {\bibfnamefont {M.}~\bibnamefont {Barbui}}, \bibinfo {author} {\bibfnamefont
  {M.~R.~D.}\ \bibnamefont {Rodrigues}}, \bibinfo {author} {\bibfnamefont
  {K.}~\bibnamefont {Schmidt}}, \bibinfo {author} {\bibfnamefont
  {D.}~\bibnamefont {Fabris}}, \bibinfo {author} {\bibfnamefont
  {M.}~\bibnamefont {Lunardon}}, \bibinfo {author} {\bibfnamefont
  {S.}~\bibnamefont {Moretto}}, \bibinfo {author} {\bibfnamefont
  {G.}~\bibnamefont {Nebbia}}, \bibinfo {author} {\bibfnamefont
  {S.}~\bibnamefont {Pesente}}, \bibinfo {author} {\bibfnamefont
  {V.}~\bibnamefont {Rizzi}}, \bibinfo {author} {\bibfnamefont
  {G.}~\bibnamefont {Viesti}}, \bibinfo {author} {\bibfnamefont
  {M.}~\bibnamefont {Cinausero}}, \bibinfo {author} {\bibfnamefont
  {G.}~\bibnamefont {Prete}}, \bibinfo {author} {\bibfnamefont
  {T.}~\bibnamefont {Keutgen}}, \bibinfo {author} {\bibfnamefont
  {Y.}~\bibnamefont {El~Masri}},\ and\ \bibinfo {author} {\bibfnamefont
  {Z.}~\bibnamefont {Majka}},\ }\bibfield  {title} {\bibinfo {title}
  {Experimental {{Determination}} of {{In-Medium Cluster Binding Energies}} and
  {{Mott Points}} in {{Nuclear Matter}}},\ }\href
  {https://doi.org/10.1103/PhysRevLett.108.062702} {\bibfield  {journal}
  {\bibinfo  {journal} {Phys. Rev. Lett.}\ }\textbf {\bibinfo {volume} {108}},\
  \bibinfo {pages} {062702} (\bibinfo {year} {2012})}\BibitemShut {NoStop}%
\bibitem [{\citenamefont {Pais}\ \emph {et~al.}(2020)\citenamefont {Pais},
  \citenamefont {Bougault}, \citenamefont {Gulminelli}, \citenamefont
  {Provid{\^e}ncia}, \citenamefont {Bonnet}, \citenamefont {Borderie},
  \citenamefont {Chbihi}, \citenamefont {Frankland}, \citenamefont {Galichet},
  \citenamefont {Gruyer}, \citenamefont {Henri}, \citenamefont {Le~Neindre},
  \citenamefont {Lopez}, \citenamefont {Manduci}, \citenamefont {Parl{\^o}g},\
  and\ \citenamefont {Verde}}]{PaiPRL125}%
  \BibitemOpen
  \bibfield  {author} {\bibinfo {author} {\bibfnamefont {H.}~\bibnamefont
  {Pais}}, \bibinfo {author} {\bibfnamefont {R.}~\bibnamefont {Bougault}},
  \bibinfo {author} {\bibfnamefont {F.}~\bibnamefont {Gulminelli}}, \bibinfo
  {author} {\bibfnamefont {C.}~\bibnamefont {Provid{\^e}ncia}}, \bibinfo
  {author} {\bibfnamefont {E.}~\bibnamefont {Bonnet}}, \bibinfo {author}
  {\bibfnamefont {B.}~\bibnamefont {Borderie}}, \bibinfo {author}
  {\bibfnamefont {A.}~\bibnamefont {Chbihi}}, \bibinfo {author} {\bibfnamefont
  {J.~D.}\ \bibnamefont {Frankland}}, \bibinfo {author} {\bibfnamefont
  {E.}~\bibnamefont {Galichet}}, \bibinfo {author} {\bibfnamefont
  {D.}~\bibnamefont {Gruyer}}, \bibinfo {author} {\bibfnamefont
  {M.}~\bibnamefont {Henri}}, \bibinfo {author} {\bibfnamefont
  {N.}~\bibnamefont {Le~Neindre}}, \bibinfo {author} {\bibfnamefont
  {O.}~\bibnamefont {Lopez}}, \bibinfo {author} {\bibfnamefont
  {L.}~\bibnamefont {Manduci}}, \bibinfo {author} {\bibfnamefont
  {M.}~\bibnamefont {Parl{\^o}g}},\ and\ \bibinfo {author} {\bibfnamefont
  {G.}~\bibnamefont {Verde}},\ }\bibfield  {title} {\bibinfo {title} {Low
  {{Density In-Medium Effects}} on {{Light Clusters}} from {{Heavy-Ion
  Data}}},\ }\href {https://doi.org/10.1103/PhysRevLett.125.012701} {\bibfield
  {journal} {\bibinfo  {journal} {Phys. Rev. Lett.}\ }\textbf {\bibinfo
  {volume} {125}},\ \bibinfo {pages} {012701} (\bibinfo {year}
  {2020})}\BibitemShut {NoStop}%
\bibitem [{\citenamefont {Sumiyoshi}\ and\ \citenamefont
  {R{\"o}pke}(2008)}]{SumPRC77}%
  \BibitemOpen
  \bibfield  {author} {\bibinfo {author} {\bibfnamefont {K.}~\bibnamefont
  {Sumiyoshi}}\ and\ \bibinfo {author} {\bibfnamefont {G.}~\bibnamefont
  {R{\"o}pke}},\ }\bibfield  {title} {\bibinfo {title} {Appearance of light
  clusters in post-bounce evolution of core-collapse supernovae},\ }\href
  {https://doi.org/10.1103/PhysRevC.77.055804} {\bibfield  {journal} {\bibinfo
  {journal} {Phys. Rev. C}\ }\textbf {\bibinfo {volume} {77}},\ \bibinfo
  {pages} {055804} (\bibinfo {year} {2008})}\BibitemShut {NoStop}%
\bibitem [{\citenamefont {Oertel}\ \emph {et~al.}(2017)\citenamefont {Oertel},
  \citenamefont {Hempel}, \citenamefont {Kl{\"a}hn},\ and\ \citenamefont
  {Typel}}]{OerRMP89}%
  \BibitemOpen
  \bibfield  {author} {\bibinfo {author} {\bibfnamefont {M.}~\bibnamefont
  {Oertel}}, \bibinfo {author} {\bibfnamefont {M.}~\bibnamefont {Hempel}},
  \bibinfo {author} {\bibfnamefont {T.}~\bibnamefont {Kl{\"a}hn}},\ and\
  \bibinfo {author} {\bibfnamefont {S.}~\bibnamefont {Typel}},\ }\bibfield
  {title} {\bibinfo {title} {Equations of state for supernovae and compact
  stars},\ }\href {https://doi.org/10.1103/RevModPhys.89.015007} {\bibfield
  {journal} {\bibinfo  {journal} {Rev. Mod. Phys.}\ }\textbf {\bibinfo {volume}
  {89}},\ \bibinfo {pages} {015007} (\bibinfo {year} {2017})}\BibitemShut
  {NoStop}%
\bibitem [{\citenamefont {Danielewicz}\ and\ \citenamefont
  {Bertsch}(1991)}]{DanNPA533}%
  \BibitemOpen
  \bibfield  {author} {\bibinfo {author} {\bibfnamefont {P.}~\bibnamefont
  {Danielewicz}}\ and\ \bibinfo {author} {\bibfnamefont {G.~F.}\ \bibnamefont
  {Bertsch}},\ }\bibfield  {title} {\bibinfo {title} {Production of deuterons
  and pions in a transport model of energetic heavy-ion reactions},\ }\href
  {https://doi.org/10.1016/0375-9474(91)90541-D} {\bibfield  {journal}
  {\bibinfo  {journal} {Nucl. Phys. A}\ }\textbf {\bibinfo {volume} {533}},\
  \bibinfo {pages} {712} (\bibinfo {year} {1991})}\BibitemShut {NoStop}%
\bibitem [{\citenamefont {Danielewicz}\ and\ \citenamefont
  {Pan}(1992)}]{DanPRC46}%
  \BibitemOpen
  \bibfield  {author} {\bibinfo {author} {\bibfnamefont {P.}~\bibnamefont
  {Danielewicz}}\ and\ \bibinfo {author} {\bibfnamefont {Q.}~\bibnamefont
  {Pan}},\ }\bibfield  {title} {\bibinfo {title} {Blast of light fragments from
  central heavy-ion collisions},\ }\href
  {https://doi.org/10.1103/PhysRevC.46.2002} {\bibfield  {journal} {\bibinfo
  {journal} {Phys. Rev. C}\ }\textbf {\bibinfo {volume} {46}},\ \bibinfo
  {pages} {2002} (\bibinfo {year} {1992})}\BibitemShut {NoStop}%
\bibitem [{\citenamefont {Kuhrts}\ \emph {et~al.}(2001)\citenamefont {Kuhrts},
  \citenamefont {Beyer}, \citenamefont {Danielewicz},\ and\ \citenamefont
  {R{\"o}pke}}]{KuhPRC63}%
  \BibitemOpen
  \bibfield  {author} {\bibinfo {author} {\bibfnamefont {C.}~\bibnamefont
  {Kuhrts}}, \bibinfo {author} {\bibfnamefont {M.}~\bibnamefont {Beyer}},
  \bibinfo {author} {\bibfnamefont {P.}~\bibnamefont {Danielewicz}},\ and\
  \bibinfo {author} {\bibfnamefont {G.}~\bibnamefont {R{\"o}pke}},\ }\bibfield
  {title} {\bibinfo {title} {Medium corrections in the formation of light
  charged particles in heavy ion reactions},\ }\href
  {https://doi.org/10.1103/PhysRevC.63.034605} {\bibfield  {journal} {\bibinfo
  {journal} {Phys. Rev. C}\ }\textbf {\bibinfo {volume} {63}},\ \bibinfo
  {pages} {034605} (\bibinfo {year} {2001})}\BibitemShut {NoStop}%
\bibitem [{\citenamefont {{FOPI Collaboration}}(2010)}]{ReiNPA848}%
  \BibitemOpen
  \bibfield  {author} {\bibinfo {author} {\bibnamefont {{FOPI
  Collaboration}}},\ }\bibfield  {title} {\bibinfo {title} {Systematics of
  central heavy ion collisions in the {{$1A$ GeV}} regime},\ }\href
  {https://doi.org/10.1016/j.nuclphysa.2010.09.008} {\bibfield  {journal}
  {\bibinfo  {journal} {Nucl. Phys. A}\ }\textbf {\bibinfo {volume} {848}},\
  \bibinfo {pages} {366} (\bibinfo {year} {2010})}\BibitemShut {NoStop}%
\bibitem [{\citenamefont {Bougault}\ \emph {et~al.}(2021)\citenamefont
  {Bougault}, \citenamefont {Borderie}, \citenamefont {Chbihi}, \citenamefont
  {Fable}, \citenamefont {Frankland}, \citenamefont {Galichet}, \citenamefont
  {Genard}, \citenamefont {Gruyer}, \citenamefont {Henri}, \citenamefont
  {La~Commara}, \citenamefont {Le~Neindre}, \citenamefont {Lombardo},
  \citenamefont {Lopez}, \citenamefont {P{\^a}rlog}, \citenamefont
  {Paw{\l}owski}, \citenamefont {Verde}, \citenamefont {Vient},\ and\
  \citenamefont {Vigilante}}]{BouSmtr13}%
  \BibitemOpen
  \bibfield  {author} {\bibinfo {author} {\bibfnamefont {R.}~\bibnamefont
  {Bougault}}, \bibinfo {author} {\bibfnamefont {B.}~\bibnamefont {Borderie}},
  \bibinfo {author} {\bibfnamefont {A.}~\bibnamefont {Chbihi}}, \bibinfo
  {author} {\bibfnamefont {Q.}~\bibnamefont {Fable}}, \bibinfo {author}
  {\bibfnamefont {J.~D.}\ \bibnamefont {Frankland}}, \bibinfo {author}
  {\bibfnamefont {E.}~\bibnamefont {Galichet}}, \bibinfo {author}
  {\bibfnamefont {T.}~\bibnamefont {Genard}}, \bibinfo {author} {\bibfnamefont
  {D.}~\bibnamefont {Gruyer}}, \bibinfo {author} {\bibfnamefont
  {M.}~\bibnamefont {Henri}}, \bibinfo {author} {\bibfnamefont
  {M.}~\bibnamefont {La~Commara}}, \bibinfo {author} {\bibfnamefont
  {N.}~\bibnamefont {Le~Neindre}}, \bibinfo {author} {\bibfnamefont
  {I.}~\bibnamefont {Lombardo}}, \bibinfo {author} {\bibfnamefont
  {O.}~\bibnamefont {Lopez}}, \bibinfo {author} {\bibfnamefont
  {M.}~\bibnamefont {P{\^a}rlog}}, \bibinfo {author} {\bibfnamefont
  {P.}~\bibnamefont {Paw{\l}owski}}, \bibinfo {author} {\bibfnamefont
  {G.}~\bibnamefont {Verde}}, \bibinfo {author} {\bibfnamefont
  {E.}~\bibnamefont {Vient}},\ and\ \bibinfo {author} {\bibfnamefont
  {M.}~\bibnamefont {Vigilante}},\ }\bibfield  {title} {\bibinfo {title} {Light
  {{Cluster Production}} in {{Central Symmetric Heavy-Ion Reactions}} from
  {{Fermi}} to {{Gev Energies}}},\ }\href {https://doi.org/10.3390/sym13081406}
  {\bibfield  {journal} {\bibinfo  {journal} {Symmetry}\ }\textbf {\bibinfo
  {volume} {13}},\ \bibinfo {pages} {1406} (\bibinfo {year}
  {2021})}\BibitemShut {NoStop}%
\bibitem [{\citenamefont {R{\"o}pke}\ \emph {et~al.}(1982)\citenamefont
  {R{\"o}pke}, \citenamefont {M{\"u}nchow},\ and\ \citenamefont
  {Schulz}}]{RopNPA379}%
  \BibitemOpen
  \bibfield  {author} {\bibinfo {author} {\bibfnamefont {G.}~\bibnamefont
  {R{\"o}pke}}, \bibinfo {author} {\bibfnamefont {L.}~\bibnamefont
  {M{\"u}nchow}},\ and\ \bibinfo {author} {\bibfnamefont {H.}~\bibnamefont
  {Schulz}},\ }\bibfield  {title} {\bibinfo {title} {Particle clustering and
  {{Mott}} transitions in nuclear matter at finite temperature: ({{I}}).
  {{Method}} and general aspects},\ }\href
  {https://doi.org/10.1016/0375-9474(82)90013-6} {\bibfield  {journal}
  {\bibinfo  {journal} {Nucl. Phys. A}\ }\textbf {\bibinfo {volume} {379}},\
  \bibinfo {pages} {536} (\bibinfo {year} {1982})}\BibitemShut {NoStop}%
\bibitem [{\citenamefont {R{\"o}pke}\ \emph {et~al.}(1983)\citenamefont
  {R{\"o}pke}, \citenamefont {Schmidt}, \citenamefont {M{\"u}nchow},\ and\
  \citenamefont {Schulz}}]{RopNPA399}%
  \BibitemOpen
  \bibfield  {author} {\bibinfo {author} {\bibfnamefont {G.}~\bibnamefont
  {R{\"o}pke}}, \bibinfo {author} {\bibfnamefont {M.}~\bibnamefont {Schmidt}},
  \bibinfo {author} {\bibfnamefont {L.}~\bibnamefont {M{\"u}nchow}},\ and\
  \bibinfo {author} {\bibfnamefont {H.}~\bibnamefont {Schulz}},\ }\bibfield
  {title} {\bibinfo {title} {Particle clustering and {{Mott}} transition in
  nuclear matter at finite temperature ({{II}}): {{Self-consistent}} ladder
  {{Hartree-Fock}} approximation and model calculations for cluster abundances
  and the phase diagram},\ }\href
  {https://doi.org/10.1016/0375-9474(83)90265-8} {\bibfield  {journal}
  {\bibinfo  {journal} {Nucl. Phys. A}\ }\textbf {\bibinfo {volume} {399}},\
  \bibinfo {pages} {587} (\bibinfo {year} {1983})}\BibitemShut {NoStop}%
\bibitem [{\citenamefont {Rammer}(2007)}]{Ram2007}%
  \BibitemOpen
  \bibfield  {author} {\bibinfo {author} {\bibfnamefont {J.}~\bibnamefont
  {Rammer}},\ }\href@noop {} {\emph {\bibinfo {title} {Quantam {{Field Theory}}
  of {{Non-Equilibrium States}}}}}\ (\bibinfo  {publisher} {{Cambridge
  University Press}},\ \bibinfo {address} {{New York}},\ \bibinfo {year}
  {2007})\BibitemShut {NoStop}%
\bibitem [{\citenamefont {Bertsch}\ and\ \citenamefont
  {Das~Gupta}(1988)}]{BerPR160}%
  \BibitemOpen
  \bibfield  {author} {\bibinfo {author} {\bibfnamefont {G.~F.}\ \bibnamefont
  {Bertsch}}\ and\ \bibinfo {author} {\bibfnamefont {S.}~\bibnamefont
  {Das~Gupta}},\ }\bibfield  {title} {\bibinfo {title} {A guide to microscopic
  models for intermediate energy heavy ion collisions},\ }\href
  {https://doi.org/10.1016/0370-1573(88)90170-6} {\bibfield  {journal}
  {\bibinfo  {journal} {Phys. Rep.}\ }\textbf {\bibinfo {volume} {160}},\
  \bibinfo {pages} {189} (\bibinfo {year} {1988})}\BibitemShut {NoStop}%
\bibitem [{\citenamefont {{TMEP collaboration}}(2022)}]{WolPPNP125}%
  \BibitemOpen
  \bibfield  {author} {\bibinfo {author} {\bibnamefont {{TMEP
  collaboration}}},\ }\bibfield  {title} {\bibinfo {title} {Transport model
  comparison studies of intermediate-energy heavy-ion collisions},\ }\href
  {https://doi.org/10.1016/j.ppnp.2022.103962} {\bibfield  {journal} {\bibinfo
  {journal} {Prog. Part. Nucl. Phys.}\ }\textbf {\bibinfo {volume} {125}},\
  \bibinfo {pages} {103962} (\bibinfo {year} {2022})}\BibitemShut {NoStop}%
\bibitem [{\citenamefont {Carlson}\ \emph {et~al.}(1973)\citenamefont
  {Carlson}, \citenamefont {Doherty}, \citenamefont {Margaziotis},
  \citenamefont {Slaus}, \citenamefont {Tin},\ and\ \citenamefont {{van
  Oers}}}]{CarLNC8}%
  \BibitemOpen
  \bibfield  {author} {\bibinfo {author} {\bibfnamefont {R.~F.}\ \bibnamefont
  {Carlson}}, \bibinfo {author} {\bibfnamefont {P.}~\bibnamefont {Doherty}},
  \bibinfo {author} {\bibfnamefont {D.~J.}\ \bibnamefont {Margaziotis}},
  \bibinfo {author} {\bibfnamefont {I.}~\bibnamefont {Slaus}}, \bibinfo
  {author} {\bibfnamefont {S.~Y.}\ \bibnamefont {Tin}},\ and\ \bibinfo {author}
  {\bibfnamefont {W.~T.~H.}\ \bibnamefont {{van Oers}}},\ }\bibfield  {title}
  {\bibinfo {title} {Proton-deuteron total reaction cross-sections in the
  energy range (20{$\div$}50) {{MeV}}},\ }\href
  {https://doi.org/10.1007/BF02724711} {\bibfield  {journal} {\bibinfo
  {journal} {Lett. Nuovo Cimento}\ }\textbf {\bibinfo {volume} {8}},\ \bibinfo
  {pages} {319} (\bibinfo {year} {1973})}\BibitemShut {NoStop}%
\bibitem [{\citenamefont {Sourkes}\ \emph {et~al.}(1976)\citenamefont
  {Sourkes}, \citenamefont {Houdayer}, \citenamefont {{van Oers}},
  \citenamefont {Carlson},\ and\ \citenamefont {Brown}}]{SouPRC13}%
  \BibitemOpen
  \bibfield  {author} {\bibinfo {author} {\bibfnamefont {A.~M.}\ \bibnamefont
  {Sourkes}}, \bibinfo {author} {\bibfnamefont {A.}~\bibnamefont {Houdayer}},
  \bibinfo {author} {\bibfnamefont {W.~T.~H.}\ \bibnamefont {{van Oers}}},
  \bibinfo {author} {\bibfnamefont {R.~F.}\ \bibnamefont {Carlson}},\ and\
  \bibinfo {author} {\bibfnamefont {R.~E.}\ \bibnamefont {Brown}},\ }\bibfield
  {title} {\bibinfo {title} {Total reaction cross section for protons on
  $^{3}\mathrm{He}$ and $^4\mathrm{He}$ between 18 and 48 {{MeV}}},\ }\href
  {https://doi.org/10.1103/PhysRevC.13.451} {\bibfield  {journal} {\bibinfo
  {journal} {Phys. Rev. C}\ }\textbf {\bibinfo {volume} {13}},\ \bibinfo
  {pages} {451} (\bibinfo {year} {1976})}\BibitemShut {NoStop}%
\bibitem [{\citenamefont {Bame}\ and\ \citenamefont {Perry}(1957)}]{BamPR107}%
  \BibitemOpen
  \bibfield  {author} {\bibinfo {author} {\bibfnamefont {S.~J.}\ \bibnamefont
  {Bame}}\ and\ \bibinfo {author} {\bibfnamefont {J.~E.}\ \bibnamefont
  {Perry}},\ }\bibfield  {title} {\bibinfo {title} {$\mathrm{T}(d,
  n)\mathrm{He}^4$ {{Reaction}}},\ }\href
  {https://doi.org/10.1103/PhysRev.107.1616} {\bibfield  {journal} {\bibinfo
  {journal} {Phys. Rev.}\ }\textbf {\bibinfo {volume} {107}},\ \bibinfo {pages}
  {1616} (\bibinfo {year} {1957})}\BibitemShut {NoStop}%
\bibitem [{\citenamefont {Wong}(1982)}]{WonPRC25}%
  \BibitemOpen
  \bibfield  {author} {\bibinfo {author} {\bibfnamefont {C.-Y.}\ \bibnamefont
  {Wong}},\ }\bibfield  {title} {\bibinfo {title} {Dynamics of nuclear fluid.
  {{VIII}}. {{Time-dependent Hartree-Fock}} approximation from a classical
  point of view},\ }\href {https://doi.org/10.1103/PhysRevC.25.1460} {\bibfield
   {journal} {\bibinfo  {journal} {Phys. Rev. C}\ }\textbf {\bibinfo {volume}
  {25}},\ \bibinfo {pages} {1460} (\bibinfo {year} {1982})}\BibitemShut
  {NoStop}%
\bibitem [{\citenamefont {Lenk}\ and\ \citenamefont
  {Pandharipande}(1989)}]{LenPRC39}%
  \BibitemOpen
  \bibfield  {author} {\bibinfo {author} {\bibfnamefont {R.~J.}\ \bibnamefont
  {Lenk}}\ and\ \bibinfo {author} {\bibfnamefont {V.~R.}\ \bibnamefont
  {Pandharipande}},\ }\bibfield  {title} {\bibinfo {title} {Nuclear mean field
  dynamics in the lattice {{Hamiltonian Vlasov}} method},\ }\href
  {https://doi.org/10.1103/PhysRevC.39.2242} {\bibfield  {journal} {\bibinfo
  {journal} {Phys. Rev. C}\ }\textbf {\bibinfo {volume} {39}},\ \bibinfo
  {pages} {2242} (\bibinfo {year} {1989})}\BibitemShut {NoStop}%
\bibitem [{\citenamefont {Wang}\ \emph {et~al.}(2019)\citenamefont {Wang},
  \citenamefont {{Lie-Wen Chen}},\ and\ \citenamefont {Zhang}}]{WRPRC99}%
  \BibitemOpen
  \bibfield  {author} {\bibinfo {author} {\bibfnamefont {R.}~\bibnamefont
  {Wang}}, \bibinfo {author} {\bibnamefont {{Lie-Wen Chen}}},\ and\ \bibinfo
  {author} {\bibfnamefont {Z.}~\bibnamefont {Zhang}},\ }\bibfield  {title}
  {\bibinfo {title} {Nuclear collective dynamics in the lattice {{Hamiltonian
  Vlasov}} method},\ }\href {https://doi.org/10.1103/PhysRevC.99.044609}
  {\bibfield  {journal} {\bibinfo  {journal} {Phys. Rev. C}\ }\textbf {\bibinfo
  {volume} {99}},\ \bibinfo {pages} {044609} (\bibinfo {year}
  {2019})}\BibitemShut {NoStop}%
\bibitem [{\citenamefont {Raimondi}\ \emph {et~al.}(2011)\citenamefont
  {Raimondi}, \citenamefont {Carlsson},\ and\ \citenamefont
  {Dobaczewski}}]{RaiPRC83}%
  \BibitemOpen
  \bibfield  {author} {\bibinfo {author} {\bibfnamefont {F.}~\bibnamefont
  {Raimondi}}, \bibinfo {author} {\bibfnamefont {B.~G.}\ \bibnamefont
  {Carlsson}},\ and\ \bibinfo {author} {\bibfnamefont {J.}~\bibnamefont
  {Dobaczewski}},\ }\bibfield  {title} {\bibinfo {title} {Effective
  pseudopotential for energy density functionals with higher-order
  derivatives},\ }\href {https://doi.org/10.1103/PhysRevC.83.054311} {\bibfield
   {journal} {\bibinfo  {journal} {Phys. Rev. C}\ }\textbf {\bibinfo {volume}
  {83}},\ \bibinfo {pages} {054311} (\bibinfo {year} {2011})}\BibitemShut
  {NoStop}%
\bibitem [{\citenamefont {Wang}\ \emph {et~al.}(2018)\citenamefont {Wang},
  \citenamefont {Chen},\ and\ \citenamefont {Zhou}}]{WRPRC98}%
  \BibitemOpen
  \bibfield  {author} {\bibinfo {author} {\bibfnamefont {R.}~\bibnamefont
  {Wang}}, \bibinfo {author} {\bibfnamefont {L.-W.}\ \bibnamefont {Chen}},\
  and\ \bibinfo {author} {\bibfnamefont {Y.}~\bibnamefont {Zhou}},\ }\bibfield
  {title} {\bibinfo {title} {Extended {{Skyrme}} interactions for transport
  model simulations of heavy-ion collisions},\ }\href
  {https://doi.org/10.1103/PhysRevC.98.054618} {\bibfield  {journal} {\bibinfo
  {journal} {Phys. Rev. C}\ }\textbf {\bibinfo {volume} {98}},\ \bibinfo
  {pages} {054618} (\bibinfo {year} {2018})}\BibitemShut {NoStop}%
\bibitem [{\citenamefont {Wang}\ \emph
  {et~al.}(2020{\natexlab{a}})\citenamefont {Wang}, \citenamefont {Zhang},
  \citenamefont {Chen}, \citenamefont {Ko},\ and\ \citenamefont
  {Ma}}]{WRPLB807}%
  \BibitemOpen
  \bibfield  {author} {\bibinfo {author} {\bibfnamefont {R.}~\bibnamefont
  {Wang}}, \bibinfo {author} {\bibfnamefont {Z.}~\bibnamefont {Zhang}},
  \bibinfo {author} {\bibfnamefont {L.-W.}\ \bibnamefont {Chen}}, \bibinfo
  {author} {\bibfnamefont {C.~M.}\ \bibnamefont {Ko}},\ and\ \bibinfo {author}
  {\bibfnamefont {Y.-G.}\ \bibnamefont {Ma}},\ }\bibfield  {title} {\bibinfo
  {title} {Constraining the in-medium nucleon-nucleon cross section from the
  width of nuclear giant dipole resonance},\ }\href
  {https://doi.org/10.1016/j.physletb.2020.135532} {\bibfield  {journal}
  {\bibinfo  {journal} {Phys. Lett. B}\ }\textbf {\bibinfo {volume} {807}},\
  \bibinfo {pages} {135532} (\bibinfo {year} {2020}{\natexlab{a}})}\BibitemShut
  {NoStop}%
\bibitem [{\citenamefont {Ono}\ \emph {et~al.}(2019)\citenamefont {Ono},
  \citenamefont {Xu}, \citenamefont {Colonna}, \citenamefont {Danielewicz},
  \citenamefont {Ko}, \citenamefont {Tsang}, \citenamefont {Wang},
  \citenamefont {Wolter}, \citenamefont {Zhang}, \citenamefont {Chen},
  \citenamefont {Cozma}, \citenamefont {Elfner}, \citenamefont {Feng},
  \citenamefont {Ikeno}, \citenamefont {Li}, \citenamefont {Mallik},
  \citenamefont {Nara}, \citenamefont {Ogawa}, \citenamefont {Ohnishi},
  \citenamefont {Oliinychenko}, \citenamefont {Su}, \citenamefont {Song},
  \citenamefont {Zhang},\ and\ \citenamefont {Zhang}}]{OnoPRC100}%
  \BibitemOpen
  \bibfield  {author} {\bibinfo {author} {\bibfnamefont {A.}~\bibnamefont
  {Ono}}, \bibinfo {author} {\bibfnamefont {J.}~\bibnamefont {Xu}}, \bibinfo
  {author} {\bibfnamefont {M.}~\bibnamefont {Colonna}}, \bibinfo {author}
  {\bibfnamefont {P.}~\bibnamefont {Danielewicz}}, \bibinfo {author}
  {\bibfnamefont {C.~M.}\ \bibnamefont {Ko}}, \bibinfo {author} {\bibfnamefont
  {M.~B.}\ \bibnamefont {Tsang}}, \bibinfo {author} {\bibfnamefont {Y.-J.}\
  \bibnamefont {Wang}}, \bibinfo {author} {\bibfnamefont {H.}~\bibnamefont
  {Wolter}}, \bibinfo {author} {\bibfnamefont {Y.-X.}\ \bibnamefont {Zhang}},
  \bibinfo {author} {\bibfnamefont {L.-W.}\ \bibnamefont {Chen}}, \bibinfo
  {author} {\bibfnamefont {D.}~\bibnamefont {Cozma}}, \bibinfo {author}
  {\bibfnamefont {H.}~\bibnamefont {Elfner}}, \bibinfo {author} {\bibfnamefont
  {Z.-Q.}\ \bibnamefont {Feng}}, \bibinfo {author} {\bibfnamefont
  {N.}~\bibnamefont {Ikeno}}, \bibinfo {author} {\bibfnamefont {B.-A.}\
  \bibnamefont {Li}}, \bibinfo {author} {\bibfnamefont {S.}~\bibnamefont
  {Mallik}}, \bibinfo {author} {\bibfnamefont {Y.}~\bibnamefont {Nara}},
  \bibinfo {author} {\bibfnamefont {T.}~\bibnamefont {Ogawa}}, \bibinfo
  {author} {\bibfnamefont {A.}~\bibnamefont {Ohnishi}}, \bibinfo {author}
  {\bibfnamefont {D.}~\bibnamefont {Oliinychenko}}, \bibinfo {author}
  {\bibfnamefont {J.}~\bibnamefont {Su}}, \bibinfo {author} {\bibfnamefont
  {T.}~\bibnamefont {Song}}, \bibinfo {author} {\bibfnamefont {F.-S.}\
  \bibnamefont {Zhang}},\ and\ \bibinfo {author} {\bibfnamefont
  {Z.}~\bibnamefont {Zhang}},\ }\bibfield  {title} {\bibinfo {title}
  {Comparison of heavy-ion transport simulations: {{Collision}} integral with
  pions and {$\Delta$} resonances in a box},\ }\href
  {https://doi.org/10.1103/PhysRevC.100.044617} {\bibfield  {journal} {\bibinfo
   {journal} {Phys. Rev. C}\ }\textbf {\bibinfo {volume} {100}},\ \bibinfo
  {pages} {044617} (\bibinfo {year} {2019})}\BibitemShut {NoStop}%
\bibitem [{\citenamefont {Oliinychenko}\ \emph {et~al.}(2019)\citenamefont
  {Oliinychenko}, \citenamefont {Pang}, \citenamefont {Elfner},\ and\
  \citenamefont {Koch}}]{OliPRC99}%
  \BibitemOpen
  \bibfield  {author} {\bibinfo {author} {\bibfnamefont {D.}~\bibnamefont
  {Oliinychenko}}, \bibinfo {author} {\bibfnamefont {L.-G.}\ \bibnamefont
  {Pang}}, \bibinfo {author} {\bibfnamefont {H.}~\bibnamefont {Elfner}},\ and\
  \bibinfo {author} {\bibfnamefont {V.}~\bibnamefont {Koch}},\ }\bibfield
  {title} {\bibinfo {title} {{Microscopic study of deuteron production in PbPb
  collisions at $\sqrt{s}$ $=$ $2.76~\rm TeV$ via hydrodynamics and a hadronic
  afterburner}},\ }\href {https://doi.org/10.1103/PhysRevC.99.044907}
  {\bibfield  {journal} {\bibinfo  {journal} {Phys. Rev. C}\ }\textbf {\bibinfo
  {volume} {99}},\ \bibinfo {pages} {044907} (\bibinfo {year}
  {2019})}\BibitemShut {NoStop}%
\bibitem [{\citenamefont {Sun}\ \emph {et~al.}(2022)\citenamefont {Sun},
  \citenamefont {Wang}, \citenamefont {Ko}, \citenamefont {Ma},\ and\
  \citenamefont {Shen}}]{SKJX2022}%
  \BibitemOpen
  \bibfield  {author} {\bibinfo {author} {\bibfnamefont {K.-J.}\ \bibnamefont
  {Sun}}, \bibinfo {author} {\bibfnamefont {R.}~\bibnamefont {Wang}}, \bibinfo
  {author} {\bibfnamefont {C.~M.}\ \bibnamefont {Ko}}, \bibinfo {author}
  {\bibfnamefont {Y.-G.}\ \bibnamefont {Ma}},\ and\ \bibinfo {author}
  {\bibfnamefont {C.}~\bibnamefont {Shen}},\ }\bibfield  {title} {\bibinfo
  {title} {Unveiling the dynamics of nucleosynthesis in relativistic heavy-ion
  collisions},\ }\href {https://doi.org/10.48550/arXiv.2207.12532} {\bibfield
  {journal} {\bibinfo  {journal} {arXiv}\ ,\ \bibinfo {pages}
  {arXiv:2207.12532}} (\bibinfo {year} {2022})}\BibitemShut {NoStop}%
\bibitem [{\citenamefont {Typel}\ \emph {et~al.}(2010)\citenamefont {Typel},
  \citenamefont {R{\"o}pke}, \citenamefont {Kl{\"a}hn}, \citenamefont
  {Blaschke},\ and\ \citenamefont {Wolter}}]{TypPRC81}%
  \BibitemOpen
  \bibfield  {author} {\bibinfo {author} {\bibfnamefont {S.}~\bibnamefont
  {Typel}}, \bibinfo {author} {\bibfnamefont {G.}~\bibnamefont {R{\"o}pke}},
  \bibinfo {author} {\bibfnamefont {T.}~\bibnamefont {Kl{\"a}hn}}, \bibinfo
  {author} {\bibfnamefont {D.}~\bibnamefont {Blaschke}},\ and\ \bibinfo
  {author} {\bibfnamefont {H.~H.}\ \bibnamefont {Wolter}},\ }\bibfield  {title}
  {\bibinfo {title} {Composition and thermodynamics of nuclear matter with
  light clusters},\ }\href {https://doi.org/10.1103/PhysRevC.81.015803}
  {\bibfield  {journal} {\bibinfo  {journal} {Phys. Rev. C}\ }\textbf {\bibinfo
  {volume} {81}},\ \bibinfo {pages} {015803} (\bibinfo {year}
  {2010})}\BibitemShut {NoStop}%
\bibitem [{\citenamefont {Zhang}\ and\ \citenamefont {Chen}(2017)}]{ZZWPRC95}%
  \BibitemOpen
  \bibfield  {author} {\bibinfo {author} {\bibfnamefont {Z.-W.}\ \bibnamefont
  {Zhang}}\ and\ \bibinfo {author} {\bibfnamefont {L.-W.}\ \bibnamefont
  {Chen}},\ }\bibfield  {title} {\bibinfo {title} {Low density nuclear matter
  with light clusters in a generalized nonlinear relativistic mean-field
  model},\ }\href {https://doi.org/10.1103/PhysRevC.95.064330} {\bibfield
  {journal} {\bibinfo  {journal} {Phys. Rev. C}\ }\textbf {\bibinfo {volume}
  {95}},\ \bibinfo {pages} {064330} (\bibinfo {year} {2017})}\BibitemShut
  {NoStop}%
\bibitem [{\citenamefont {{The {S$\pi$RIT} Collaboration}}(2022)}]{LeeEPJA58}%
  \BibitemOpen
  \bibfield  {author} {\bibinfo {author} {\bibnamefont {{The {S$\pi$RIT}
  Collaboration}}},\ }\bibfield  {title} {\bibinfo {title} {Isoscaling in
  central {{Sn}}+{{Sn}} collisions at 270 {{MeV}}/u},\ }\href
  {https://doi.org/10.1140/epja/s10050-022-00851-2} {\bibfield  {journal}
  {\bibinfo  {journal} {Eur. Phys. J. A}\ }\textbf {\bibinfo {volume} {58}},\
  \bibinfo {pages} {201} (\bibinfo {year} {2022})}\BibitemShut {NoStop}%
\bibitem [{\citenamefont {Tanaka}\ \emph {et~al.}(2021)\citenamefont {Tanaka},
  \citenamefont {Yang}, \citenamefont {Typel}, \citenamefont {Adachi},
  \citenamefont {Bai}, \citenamefont {van Beek}, \citenamefont {Beaumel},
  \citenamefont {Fujikawa}, \citenamefont {Han}, \citenamefont {Heil},
  \citenamefont {Huang}, \citenamefont {Inoue}, \citenamefont {Jiang},
  \citenamefont {Kn{\"o}sel}, \citenamefont {Kobayashi}, \citenamefont
  {Kubota}, \citenamefont {Liu}, \citenamefont {Lou}, \citenamefont {Maeda},
  \citenamefont {Matsuda}, \citenamefont {Miki}, \citenamefont {Nakamura},
  \citenamefont {Ogata}, \citenamefont {Panin}, \citenamefont {Scheit},
  \citenamefont {Schindler}, \citenamefont {Schrock}, \citenamefont {Symochko},
  \citenamefont {Tamii}, \citenamefont {Uesaka}, \citenamefont {Wagner},
  \citenamefont {Yoshida}, \citenamefont {Zenihiro},\ and\ \citenamefont
  {Aumann}}]{TanSc371}%
  \BibitemOpen
  \bibfield  {author} {\bibinfo {author} {\bibfnamefont {J.}~\bibnamefont
  {Tanaka}}, \bibinfo {author} {\bibfnamefont {Z.}~\bibnamefont {Yang}},
  \bibinfo {author} {\bibfnamefont {S.}~\bibnamefont {Typel}}, \bibinfo
  {author} {\bibfnamefont {S.}~\bibnamefont {Adachi}}, \bibinfo {author}
  {\bibfnamefont {S.}~\bibnamefont {Bai}}, \bibinfo {author} {\bibfnamefont
  {P.}~\bibnamefont {van Beek}}, \bibinfo {author} {\bibfnamefont
  {D.}~\bibnamefont {Beaumel}}, \bibinfo {author} {\bibfnamefont
  {Y.}~\bibnamefont {Fujikawa}}, \bibinfo {author} {\bibfnamefont
  {J.}~\bibnamefont {Han}}, \bibinfo {author} {\bibfnamefont {S.}~\bibnamefont
  {Heil}}, \bibinfo {author} {\bibfnamefont {S.}~\bibnamefont {Huang}},
  \bibinfo {author} {\bibfnamefont {A.}~\bibnamefont {Inoue}}, \bibinfo
  {author} {\bibfnamefont {Y.}~\bibnamefont {Jiang}}, \bibinfo {author}
  {\bibfnamefont {M.}~\bibnamefont {Kn{\"o}sel}}, \bibinfo {author}
  {\bibfnamefont {N.}~\bibnamefont {Kobayashi}}, \bibinfo {author}
  {\bibfnamefont {Y.}~\bibnamefont {Kubota}}, \bibinfo {author} {\bibfnamefont
  {W.}~\bibnamefont {Liu}}, \bibinfo {author} {\bibfnamefont {J.}~\bibnamefont
  {Lou}}, \bibinfo {author} {\bibfnamefont {Y.}~\bibnamefont {Maeda}}, \bibinfo
  {author} {\bibfnamefont {Y.}~\bibnamefont {Matsuda}}, \bibinfo {author}
  {\bibfnamefont {K.}~\bibnamefont {Miki}}, \bibinfo {author} {\bibfnamefont
  {S.}~\bibnamefont {Nakamura}}, \bibinfo {author} {\bibfnamefont
  {K.}~\bibnamefont {Ogata}}, \bibinfo {author} {\bibfnamefont
  {V.}~\bibnamefont {Panin}}, \bibinfo {author} {\bibfnamefont
  {H.}~\bibnamefont {Scheit}}, \bibinfo {author} {\bibfnamefont
  {F.}~\bibnamefont {Schindler}}, \bibinfo {author} {\bibfnamefont
  {P.}~\bibnamefont {Schrock}}, \bibinfo {author} {\bibfnamefont
  {D.}~\bibnamefont {Symochko}}, \bibinfo {author} {\bibfnamefont
  {A.}~\bibnamefont {Tamii}}, \bibinfo {author} {\bibfnamefont
  {T.}~\bibnamefont {Uesaka}}, \bibinfo {author} {\bibfnamefont
  {V.}~\bibnamefont {Wagner}}, \bibinfo {author} {\bibfnamefont
  {K.}~\bibnamefont {Yoshida}}, \bibinfo {author} {\bibfnamefont
  {J.}~\bibnamefont {Zenihiro}},\ and\ \bibinfo {author} {\bibfnamefont
  {T.}~\bibnamefont {Aumann}},\ }\bibfield  {title} {\bibinfo {title}
  {Formation of {$\alpha$} clusters in dilute neutron-rich matter},\ }\href
  {https://doi.org/10.1126/science.abe4688} {\bibfield  {journal} {\bibinfo
  {journal} {Science}\ }\textbf {\bibinfo {volume} {371}},\ \bibinfo {pages}
  {260} (\bibinfo {year} {2021})}\BibitemShut {NoStop}%
\bibitem [{\citenamefont {Chomaz}\ \emph {et~al.}(2004)\citenamefont {Chomaz},
  \citenamefont {Colonna},\ and\ \citenamefont {Randrup}}]{ChoPR389}%
  \BibitemOpen
  \bibfield  {author} {\bibinfo {author} {\bibfnamefont {P.}~\bibnamefont
  {Chomaz}}, \bibinfo {author} {\bibfnamefont {M.}~\bibnamefont {Colonna}},\
  and\ \bibinfo {author} {\bibfnamefont {J.}~\bibnamefont {Randrup}},\
  }\bibfield  {title} {\bibinfo {title} {Nuclear spinodal fragmentation},\
  }\href {https://doi.org/10.1016/j.physrep.2003.09.006} {\bibfield  {journal}
  {\bibinfo  {journal} {Phys. Rep.}\ }\textbf {\bibinfo {volume} {389}},\
  \bibinfo {pages} {263} (\bibinfo {year} {2004})}\BibitemShut {NoStop}%
\bibitem [{\citenamefont {Wang}\ \emph
  {et~al.}(2020{\natexlab{b}})\citenamefont {Wang}, \citenamefont {Ma},
  \citenamefont {Wada}, \citenamefont {Chen}, \citenamefont {He}, \citenamefont
  {Liu},\ and\ \citenamefont {Sun}}]{WRPRR2}%
  \BibitemOpen
  \bibfield  {author} {\bibinfo {author} {\bibfnamefont {R.}~\bibnamefont
  {Wang}}, \bibinfo {author} {\bibfnamefont {Y.-G.}\ \bibnamefont {Ma}},
  \bibinfo {author} {\bibfnamefont {R.}~\bibnamefont {Wada}}, \bibinfo {author}
  {\bibfnamefont {L.-W.}\ \bibnamefont {Chen}}, \bibinfo {author}
  {\bibfnamefont {W.-B.}\ \bibnamefont {He}}, \bibinfo {author} {\bibfnamefont
  {H.-L.}\ \bibnamefont {Liu}},\ and\ \bibinfo {author} {\bibfnamefont {K.-J.}\
  \bibnamefont {Sun}},\ }\bibfield  {title} {\bibinfo {title} {Nuclear
  liquid-gas phase transition with machine learning},\ }\href
  {https://doi.org/10.1103/PhysRevResearch.2.043202} {\bibfield  {journal}
  {\bibinfo  {journal} {Phys. Rev. Research}\ }\textbf {\bibinfo {volume}
  {2}},\ \bibinfo {pages} {043202} (\bibinfo {year}
  {2020}{\natexlab{b}})}\BibitemShut {NoStop}%
\bibitem [{\citenamefont {Liu}\ \emph {et~al.}(2022)\citenamefont {Liu},
  \citenamefont {Deng},\ and\ \citenamefont {Ma}}]{LCNST33}%
  \BibitemOpen
  \bibfield  {author} {\bibinfo {author} {\bibfnamefont {C.}~\bibnamefont
  {Liu}}, \bibinfo {author} {\bibfnamefont {X.-G.}\ \bibnamefont {Deng}},\ and\
  \bibinfo {author} {\bibfnamefont {Y.-G.}\ \bibnamefont {Ma}},\ }\bibfield
  {title} {\bibinfo {title} {Density fluctuations in intermediate-energy
  heavy-ion collisions},\ }\href {https://doi.org/10.1007/s41365-022-01040-y}
  {\bibfield  {journal} {\bibinfo  {journal} {NUCL. SCI. TECH.}\ }\textbf
  {\bibinfo {volume} {33}},\ \bibinfo {pages} {52} (\bibinfo {year}
  {2022})}\BibitemShut {NoStop}%
\bibitem [{\citenamefont {Colonna}\ \emph {et~al.}(1998)\citenamefont
  {Colonna}, \citenamefont {Di~Toro}, \citenamefont {Guarnera}, \citenamefont
  {Maccarone}, \citenamefont {{Zielinska-Pfab{\'e}}},\ and\ \citenamefont
  {Wolter}}]{ColNPA642}%
  \BibitemOpen
  \bibfield  {author} {\bibinfo {author} {\bibfnamefont {M.}~\bibnamefont
  {Colonna}}, \bibinfo {author} {\bibfnamefont {M.}~\bibnamefont {Di~Toro}},
  \bibinfo {author} {\bibfnamefont {A.}~\bibnamefont {Guarnera}}, \bibinfo
  {author} {\bibfnamefont {S.}~\bibnamefont {Maccarone}}, \bibinfo {author}
  {\bibfnamefont {M.}~\bibnamefont {{Zielinska-Pfab{\'e}}}},\ and\ \bibinfo
  {author} {\bibfnamefont {H.~H.}\ \bibnamefont {Wolter}},\ }\bibfield  {title}
  {\bibinfo {title} {Fluctuations and dynamical instabilities in heavy-ion
  reactions},\ }\href {https://doi.org/10.1016/S0375-9474(98)00542-9}
  {\bibfield  {journal} {\bibinfo  {journal} {Nucl. Phys. A}\ }\textbf
  {\bibinfo {volume} {642}},\ \bibinfo {pages} {449} (\bibinfo {year}
  {1998})}\BibitemShut {NoStop}%
\end{thebibliography}
\end{document}